\documentclass[twocolumn,aps,pra,groupedaddress,superscriptaddress,amsmath,amssymb,showpacs,noeprint,english]{revtex4-1}
\usepackage[T1]{fontenc}
\usepackage[latin9]{inputenc}
\usepackage{mathrsfs}
\usepackage{amsbsy}
\usepackage{amstext}
\usepackage{amssymb}
\usepackage{setspace}
\usepackage{esint}
\usepackage{xcolor,colortbl}

\makeatletter
\usepackage{graphicx}   % need for figures

\makeatother

\usepackage{babel}
\begin{document}

\title{Non-Markovian memory in a measurement-based quantum computer}

\author{D. Filenga}
\affiliation{Faculdade de Ciências, UNESP - Universidade Estadual Paulista, 17033-360 Bauru, São Paulo, Brazil}
\author{F. Mahlow}
\affiliation{Faculdade de Ciências, UNESP - Universidade Estadual Paulista, 17033-360 Bauru, São Paulo, Brazil}
\author{F. F. Fanchini}
\affiliation{Faculdade de Ciências, UNESP - Universidade Estadual Paulista, 17033-360 Bauru, São Paulo, Brazil}

\date{\today}
%=======================================================================================
\begin{abstract}
%=======================================================================================
We study the exact open system dynamics of single qubit gates during a measurement-based quantum computation considering non-Markovian environments. We obtain analytical solutions for the average gate fidelities and analyze it for amplitude damping and dephasing channels. We show that the average fidelity is identical for the $X$-gate and $Z$-gate and that neither fast application of the projective measurements necessarily implies high gate fidelity, nor slow application necessarily implies low gate fidelity. Indeed, for highly non-Markovian environments, it is of utmost importance to know the best time to perform the measurements, since a huge variation in the gate fidelity may occur given this scenario. Furthermore, we show that while for the amplitude damping the knowledge of the dissipative map is sufficient to determine the best measurement times, i.e. the best times in which measures are taken, the same is not necessarily true for the phase damping. To the later, the time of the set of measures becomes crucial since a phase error in one qubit can fix the phase error that takes place in another. %OK
\end{abstract}
\maketitle
%=======================================================================================
\section{Introduction}
%=======================================================================================
Quantum computing and information theory are research areas that bring great impact to society \cite{impact}. It is a topic of great interest, both academically and technologically, as it allows the processing of a large amount of information, for certain tasks, which is impractical or impossible to be performed in a classical computer \cite{nielsen}.
Over the past few years, a number of technological advances emerged and new strategies, such as measurement-based quantum computation (MBQC) \cite{raussendorf}, have been contributing to the development of a more robust quantum computer.
This technique differs from the standard quantum computation scheme, as it uses projective measurements on special entangled states instead of unitary operations and, in recent years, different experiments have been performed to demonstrate the feasibility of a MBQC \cite{experiment}.

One of the great challenges for the development of quantum computers is dealing with environmental noise. Interactions with the environment are unavoidable and decrease the computation fidelity due to the decoherence process, where the quantum properties are lost \cite{breuer}. Thereby, the study of the open quantum system dynamics, in order to better understand the dissipative processes, becomes fundamental to obtain higher computing fidelities.
In special, for non-Markovian process, the quantum state coherence presents a non-monotonic dynamic, a subject of broad interest \cite{nonm}, and a feature especially relevant to MBQC. Indeed, for a quantum system interacting with non-Markovian environments, the coherence can be re-established at certain time points \cite{bellomo}, since a measurement done in the right time can result in higher fidelity values. It is in this direction, understanding how a non-Markovian environment can influence a measurement-based quantum computer that we develop our studies. %OK

Here, we consider two different and quite common non-Markovian dissipative process: the amplitude damping (AD) \cite{breuer} and phase damping (PD) \cite{daffer}. We obtain an analytical solution for the average gate fidelities considering the initial conditions and the measurement times of the MBQC, and we show how each measurement can highly interfere in this kind of quantum computing. At this purpose, we organize this manuscript as follows: in section II we describe the MBQC technique for a simple $X$ and $Z$ gates; in section III, we describe the non-Markovian dissipative model for amplitude and phase damping noises; in section IV, we discuss the resource gate fidelity \cite{wang} and the average gate fidelity; in section V we present the implications of our results to the MBQC and, in section V, we conclude the manuscript. %OK
%=======================================================================================
\section{MBQC}
%=======================================================================================
We suppose the simpler procedure to study the MBQC considering interaction with non-Markovian environments: a linear cluster state and one qubit gate. The limitation in our choice is due to the significant difficulty in calculating the average gate fidelity, which takes into account an average over all initial conditions and all possible measurement sequences of the MBQC. The computational procedure was firstly introduced by Raussendorf and Breuer in their seminal paper \cite{raussendorf}. The idea begins with a five qubit array defined as $|\Psi(\psi_{in})\rangle=|\psi_{in}\rangle_1\otimes |+\rangle_2\otimes |+\rangle_3\otimes |+\rangle_4\otimes |+\rangle_5$, where $|\psi_{in}\rangle$ is an arbitrary qubit initial state and $|+\rangle$ is one of the eigenvectors of the Pauli matrix $\sigma_x$. In sequence, a highly entangled state, called cluster state, is generated introducing an Ising-type next-neighbor interaction between the qubits which is described by the Hamiltonian $H_{\rm int}=\sum_{j=1}^4\frac{1+\sigma_z^{j}}{2}\frac{1-\sigma_z^{j+1}}{2}$ where $j$ determines the qubit site and $\sigma_z$ is the well-known Pauli $z$ matrix. The cluster state $|\Psi_C(\psi_{in})\rangle$ is thus, created by applying $S$ on $|\Psi(\psi_{in})\rangle$, i.e. $|\Psi_C(\psi_{in})\rangle=S|\Psi(\psi_{in})\rangle$, where $S=\exp(-i\pi H_{\rm int})$. %OK

The focus of our studies is the implementation of a $\pi$ rotation in the $X$ or $Z$ direction. Indeed, as we will show here, the average gate fidelity will be identical, whether it is a NOT (also called $X$) gate or a $Z$ gate.  The state $|\psi_{in}\rangle$ can be rotated by measuring qubits 1 to 4, at the same time that the final result, i.e. the rotated state, is teleported to qubit 5. The implementation of these two gates is, indeed, equivalent. The idea is to measure all four qubit on the basis of the Pauli $\sigma_x$ operator. The resulting state is $|\psi_{out}\rangle = X^{s_2+s_4}Z^{s_1+s_3}U|\psi_{in}\rangle$, where $U$ is the desired gate ($X$ or $Z$), and $s_j \in \{0,1\}$ for $j=1,...,4$, depends on the measurement results and the desired gate. To the case where $U\equiv X$ (the $X$ gate), on the one hand, $s_2=+1$ (0) if the second qubit collapses in the $|+\rangle$ ($|-\rangle$) state, and $s_1=s_3=s_4=0$ (+1) if the first, third and fourth qubit collapse in the $|+\rangle$ ($|-\rangle$) state. On the other hand, to the case where $U\equiv Z$ (the $Z$ gate), $s_3=+1$ (0) if the third qubit collapses in the $|+\rangle$ ($|-\rangle$) state, and $s_1=s_2=s_4=0$ (+1) if the first, second and fourth qubit collapse in the $|+\rangle$ ($|-\rangle$) state. %OK

As we can note, unless an extra rotation given by $X^{s_2+s_4}Z^{s_1+s_3}$, where ${\mathbf s} = \{s_1,s_2,s_3,s_4 \}$ is the set of all possible results for $s_1$, $s_2$, $s_3$, and $s_4$, the desired gate is implemented and the intended output state emerges in the fifth qubit. This extra rotation can be understood as a basis change, in which the final answer is expressed on a basis different from that initially defined.
Indeed, this additional operation does not depend on the initial state and is developed based on the measurement results, since, under no circumstances limits the computational power of the MBQC. Thus, after correcting the resulting state by means of the $B_{\mathbf s}=Z^{s_1+s_3}X^{s_2+s_4}$ operation we get the desired gate and output state. In other words, $B_{\mathbf s}|\psi_{out}\rangle=U|\psi_{in}\rangle$. %OK
%=======================================================================================
\section{Dissipative Model}
%=======================================================================================
In this section we will describe the dissipative model used to investigate the unusual consequences of memory effects that occur in a non-Markovian environment. In this context, we model two kinds of channels: amplitude damping and phase damping (or dephasing). Both noises are dissipative processes that occur independently in each qubit, each one coupled to its respective reservoir.
To calculate the dissipative dynamics, we utilize the Kraus operators, a friendly and viable method that allows us to obtain the analytical expressions for the average fidelity \cite{nielsen}. %OK
%---------------------------------------------------------------------------------------
\subsection{Amplitude Damping} 
%---------------------------------------------------------------------------------------
The amplitude damping channel can describe the general energy dissipation behaviour of different quantum systems. For instance, it can describe the state of a photon in a cavity subject to scattering as well as the dynamics of an atom emitting a photon spontaneously \cite{nielsen}.
In this case, the environment can be represented by a bath of harmonic oscillators with spectral density defined by \cite{breuer}:
\begin{equation}
J(\omega ) = \frac{1}{{2\pi }}\frac{{{\gamma _0}{\lambda ^2}}}{{{{({\omega _0} - \omega )}^2} + {\lambda ^2}}},
\end{equation}
with $\lambda\approx1/\tau_B$, where $\tau_B$ is the reservoir correlation time and $\gamma_0$ is given by $ {\gamma _0}\approx 1/{\tau _R}$, where $\tau_R$ is a typical system time scale. Strong coupling occurs when ${\tau _R} < 2{\tau _B}$.
The set of Kraus operators used to describe one qubit dynamics can be expressed as
 \cite{nielsen}:
\begin{equation}\label{eap}
    \begin{array}{*{20}{c}}
{E_1(t) = \left( {\begin{array}{*{20}{c}}
1&0\\
0&{\sqrt {{p(t)}} }
\end{array}} \right)}&{\rm and}&{E_2(t) = \left( {\begin{array}{*{20}{c}}
0&{\sqrt {1 - {{p(t)}}} }\\
0&0
\end{array}} \right)}
\end{array},
\end{equation}
with $p(t)$ given by
\begin{equation}\label{ptad}
   p(t)= {e^{ - {\lambda}t}}{\left[ {\frac{{{\lambda}}}{{{d}}}\sin\!\left( {\frac{{{d}t}}{2}} \right) + \cos\!\left( {\frac{{{d}t}}{2}} \right)} \right]^2},
\end{equation}
where $d = \sqrt {2{\gamma_0}\lambda  - {\lambda^2}}$. %OK
%---------------------------------------------------------------------------------------
\subsection{Dephasing}
%---------------------------------------------------------------------------------------
The behaviour of a randomly dispersing photon or the perturbation of the electronic states in an atom that interacts with distant electric charges are some examples which can be described through the phase damping or dephasing channel \cite{nielsen}.
The set of Kraus operators that can be used to describe the dissipative dynamics of one qubit, when subjected to this kind of process, can be expressed by \cite{daffer}:
\begin{equation}\label{epd1}
{E_1}(t) = \left( {\begin{array}{*{20}{c}}
{\sqrt {\frac{{L(t) + 1}}{2}} }&0\\
0&{\sqrt {\frac{{L(t) + 1}}{2}} }
\end{array}} \right)
\end{equation}
and
\begin{equation}\label{epd2}
{E_2}(t) = \left( {\begin{array}{*{20}{c}}
{\sqrt {\frac{{1 - L(t)}}{2}} }&0\\
0&{ - \sqrt {\frac{{1 - L(t)}}{2}} }
\end{array}} \right),
\end{equation}
with $L(t)$ given by
\begin{equation}
    L(t) = {e^{ - t/2\tau }}\left[ {\frac{1}{u}\sin \left( {\frac{{ut}}{{2\tau }}} \right) + \cos \left( {\frac{{ut}}{{2\tau }}} \right)} \right],
	\label{ltpd}
\end{equation}
where $u = \sqrt {16{a^2}{\tau ^2} - 1}$. 

This model describes a colored noise, where the system is coupled to some preferable frequencies. In this sense, the coupling with the external system is strengthen by $a$, while $\tau$ determines which frequencies the system prefers most \cite{daffer}. %OK
%=======================================================================================
\section{Fidelity}
%=======================================================================================
To study the efficiency of the measurement based quantum computer we consider two distinct fidelities: the resource gate fidelity \cite{wang}, which measures the capacity of teleporting a gate, and the average gate fidelity, that gives the average fidelity of the computation as a function of the initial state. The latter has a direct operational interpretation since it represents the average fidelity of the computation given an arbitrary initial condition. Below we describe both in detail, explaining how they are defined considering a MBQC. %OK
%---------------------------------------------------------------------------------------
\subsection{Resource Gate Fidelity}
%---------------------------------------------------------------------------------------
The resource gate fidelity has been used to define how well a quantum gate can be implemented on a measurement-based quantum computer. Despite the facility of its implementation, since the average is just calculated over all possible measurement output, it suffers from a lack of interpretation. Indeed, the fidelity for gate operations do not represent the average fidelity obtained in a real experiment and, as we will show below, produces quite different results when compared to it. The resource gate fidelity is defined employing a process called gate teleportation \cite{gote}. The idea is based on the teleportation of a qubit by means of a rotated EPR pair, where the rotation defines some unitary operation. It is called resource state and it is defined as $(I\otimes U)(|00\rangle+|11\rangle)$ where $U$ is a desired unitary operation \cite{wang}. The previous proposed scheme to define the fidelity of a MBQC relies on the fact that a cluster state can be used to prepare resource states since this method offers a simple way to define a gate fidelity. We will focus our studies considering two different gates, the $X$ and $Z$ gate where %OK
\begin{equation}
    \begin{array}{*{20}{c}}
{X = \left( {\begin{array}{*{20}{c}}
0&1\\
1&0
\end{array}} \right)}&{\rm and}&{Z = \left( {\begin{array}{*{20}{c}}
1&0\\
0&-1
\end{array}} \right)}
\end{array}.
\end{equation}

To prepare the $X$ or $Z$ resource state by means of a five-qubit linear cluster state, we use a sequence of three measurements, applied on qubit 2, 3, and 4, with all of them applied on the basis of the Pauli $\sigma_x$ operator. The resulting state, as a function of the $Z$ resource state, is %OK
\begin{equation}
X_5^{r_4}Z_5^{r_3}X_5^{r_2}\underbrace{[I\otimes Z(|00\rangle+|11\rangle)_{15}]}_{ {\rm Z\;resource\;state}}
\end{equation}
where $r_2$, $r_4=+1$ (0) if the second and fourth qubit collapse, respectively, in the $|+\rangle$ ($|-\rangle$) state and $r_3=0$ (+1) if the third qubit collapse in the $|+\rangle$ ($|-\rangle$) state.
Thus, as usual in a measurement-based quantum computer, after correcting the resulting state by means of the $B_{\mathbf r} = X_5^{r_2} Z_5^{r_3}X_5^{r_4}$ operation, where ${\mathbf r}=\{r_1,r_2,r_3\}$ is the set of all possible results for $r_1$, $r_2$, and $r_3$, we get the desired $Z$ resource state. On the other hand, with the same sequence of measurements, with qubit 2, 3, and 4 measured on the Pauli $\sigma_x$ basis, the resulting state can also be written as a function of the $X$ resource state %OK
\begin{equation}
Z_5^{r_2+r_3+r_4}X_5^{r_2+r_4+1}Z_5^{r_2+r_4}\underbrace{[I\otimes X(|00\rangle+|11\rangle)_{15}]}_{ {\rm X\;resource\;state}}
\end{equation}
where $r_2, r_3, r_4=+1$ (0) if the second, third, and fourth qubit collapse, respectively, in the $|+\rangle$ ($|-\rangle$) state. Again, after correcting the resulting state by means of the $B_{\mathbf r} = Z_5^{r_2+r_4}X_5^{r_2+r_4+1}Z_5^{r_2+r_3+r_4}$ operation, we obtain the desired $X$ resource state.
The resource gate fidelity is defined as %OK
\begin{equation}
F_{res} = {\rm Tr}\; \rho_r |\Psi_{res}\rangle\langle\Psi_{res}|,
\end{equation}
where $|\Psi_{res}\rangle$ is the resource state, and $\rho_r$ is the resulting two-qubit state after the decoherent process, the sequence of measurements and it respective correction $B_{\mathbf r}$. In other words, %OK
\begin{equation}
\rho_r= \frac{1}{8}\;{\rm Tr}_{234}\;\sum_{\mathbf r} B_{\mathbf r}P_{\mathbf r}\mathcal L[\rho_C(+)]P_{\mathbf r}B_{\mathbf r}^\dagger
\end{equation}
where the sum is over all measuring outcome possible sequence, $\rho_C(+)$ is a cluster state with $|\psi_{in}\rangle=|+\rangle$, i.e. $\rho_C(+)=|\Psi_C(+)\rangle\langle\Psi_C(+)|$, $\mathcal L$ is the decoherent superopetaror, $P_{\mathbf r}$ is the projection operator, and $B_{\mathbf r}$ is the error-correction operator. Here, the partial trace operation traces out qubits other than the ones of the resource state, $P_{\mathbf r}$ take into account all possible measurement sequence on qubit 2, 3, and 4, while $B_{\mathbf r}$ introduces the necessary correction depending on the desired resource state, $X$ or $Z$. Finally, the sum introduces an average over all possible measurement results. It is straightforward to show that if $\mathcal L [\rho_C(+)]=\rho_C(+)$, we have $\rho_r =|\Psi_{res}\rangle\langle\Psi_{res}|$, since the fidelity turns out to be 1 in this case. %OK
%---------------------------------------------------------------------------------------
\subsection{Average Gate Fidelity}
%---------------------------------------------------------------------------------------
Although resource gate fidelity was used to determine the gate fidelity in a MBQC, their final result does not provide an operational interpretation when concerning a measurement-based quantum computation. Indeed, the most natural definition is the average gate fidelity, which takes into account the average over all possible initial conditions and sequence of measurements. In this sense, we define the average gate fidelity as %OK
\begin{equation}
F_{gate} = \frac{1}{\mathcal N}\sum_{\{\psi_{in}\}}{\rm Tr}\;\rho_{out}(\psi_{in})\rho_s(\psi_{in})\label{fgate}
\end{equation}
with $\rho_{out}(\psi_{in}) = U|\psi_{in}\rangle\langle\psi_{in}|U^\dagger$, where $U$ is the desired gate,  $\mathcal N$ is the normalization factor which can be written as $\mathcal N = \sum_{\{\psi_{in}\}}{\rm Tr}\;[\rho_{out}(\psi_{in})]$, and
\begin{equation}
\rho_s(\psi_{in})= \frac{1}{16}\;{\rm Tr}_{1234}\;\sum_{\mathbf s} B_{\mathbf s}P_{\mathbf s}\mathcal L[\rho_C(\psi_{in})]P_{\mathbf s}B_{\mathbf s}^\dagger
\end{equation}  
where the sum is over all measuring outcome possible sequence, $\rho_C(\psi_{in})=|\Psi_C(\psi_{in})\rangle\langle \Psi_C(\psi_{in})|$ is a cluster state as a function of  an arbitrary qubit initial state, $\mathcal L$ is the decoherent superopetaror, $P_{\mathbf s}$ is the projection operator, and $B_{\mathbf s}$ is the error-correction operator. Here, the partial trace operation traces out all qubits other than the output state, $P_{\mathbf s}$ takes into account all possible measurement sequence on qubit 1, 2, 3, and 4, while $B_{\mathbf s}$ introduces the necessary correction depending on the desired resource state, $X$ or $Z$. Finally, the sum on $\mathbf s$ introduces an average over all possible measurement results and the sum on $\psi_{in}$ in Eq. (\ref{fgate}) over all possible initial states. This definition gives the average fidelity reached in a real experiment where the initial state is arbitrary. %OK
%=======================================================================================
\section{Results}
\begin{figure}[h!]
\centering
\includegraphics[scale=0.3]{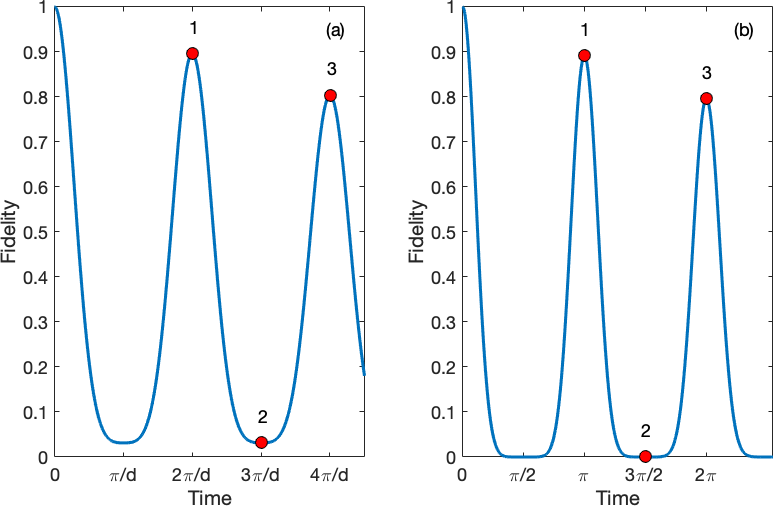}
\caption{Average fidelity of cluster dissipative dynamics as a function of time to the case of the amplitude damping channel (a) and to the case of the phase damping channel (b). The red circles enumerated with 1, 2, and 3 defines a set of preselected times where the measurement times are performed.}\label{fig1} %OK
\end{figure}

%=======================================================================================
To analyze the decoherent dynamics of a measurement-based quantum computer, we first study the decoherent dynamics of the cluster state, without considering any gate and therefore no measurements. The initial state is given by $\rho_C(\psi_{in})$ and we calculate the average fidelity over a set of 10100 initial conditions $\{\psi_{in}\}$ equally distributed on the Bloch sphere. Thus, %OK
\begin{equation}
F(t)=\frac{1}{\mathcal N}\sum_{\{\psi_{in}\}}{{\rm Tr}\; \mathcal{L}_t[\rho_C(\psi_{in})]\;\rho_C(\psi_{in})}
\end{equation}
where the normalization factor is given by $\mathcal N = \sum_{\{\psi_{in}\}}{\rm Tr}\;[\rho_C(\psi_{in})]$ and 
\begin{equation}
 \mathcal{L}_t[\rho_C(\psi_{in})]=\sum_{i,j,k,l,m=1}^2 M_{ijklm}(t)\;\rho_C(\psi_{in})\;M_{ijklm}^\dagger(t)
 \end{equation}
where $M_{ijklm}(t)=E_i^{(1)}(t)E_j^{(2)}(t)E_k^{(3)}(t)E_l^{(4)}(t)E_m^{(5)}(t)$ with $E_q^{(n)}$ given by the q-\textit{th} Kraus operator acting on the n-\textit{th} qubit. Note that depending on the decoherent process, amplitude damping or phase damping, the set of Kraus operators are given by Eq.(\ref{eap}) or Eq.(\ref{epd1}) and Eq.(\ref{epd2}), respectively. Here we consider identical and independent environment, since each qubit has it own reservoir. Also, we consider highly non-Markovian environments since we set $\lambda=10^{-3}$ and $\gamma_0=10$ to define the amplitude damping channel in Eq. (\ref{ptad}), and $a=1$ and $\tau_0=30$ to define the phase damping channel in Eq. (\ref{ltpd}). In Fig. (\ref{fig1}-a) we show the average fidelity as a function of time to the case of the amplitude damping channel and in Fig. (\ref{fig1}-b) to the case of the phase damping. The main purpose of the dissipative dynamics analysis of cluster states is to study the fidelities when the measurements, responsible for the logical operations, are made in their peaks and valleys. %OK

As we can note, the high degree of non-Markovianity introduces a highly non-monotonical behaviour of the cluster state fidelity. Thus, considering this scenario, how is the gate fidelity of a MBQC influenced by the measurement times? Naturally, if all measurements are applied at times close to $t=0$, the fidelity is approximately equal to 1, but what could we say if a small delay occurs in the first measurement?
To answer this question, we study the MBQC gate fidelity as a function of the measurement times, i.e. considering all possible time combinations within a set of preselected times. The analytical expressions for the average gate fidelities, as a function of the measurement times, can be found on the supplemental material at [URL will be inserted by publisher].
We consider three distinct set of times for the measurements, $2\pi/d$, $3\pi/d$, and $4\pi/d$ to the amplitude damping case (see the red circles in Fig. (\ref {fig1}-a)) and $\pi$, $3\pi/2$, and $2\pi$ to the phase damping case (see the red circles in Fig. (\ref{fig1}-b)). As we can note in Fig. (\ref{fig1}) these times are related to a peak (1), a valley (2), and another peak (3) of the cluster state fidelity.
The resulting fidelities are presented in Table \ref{tab1} to the average gate fidelity and in Table \ref{tab2} to the resource gate fidelity. %OK

\def\arraystretch{1.5}%
%\begin{table}[h!]
 {\centering
 \begin{table}
    \begin{tabular}{|c|c|c|}
    \hline
      \textbf{Measurements} & \textbf{Fidelity of the} & \textbf{Fidelity of the} \\ 
       \textbf{$\;${Time}$\;$} & \textbf{AD Channel} & \textbf{PD Channel} \\ 
      \hline
      1-1-1-1 & 0.958 & 0.928 \\\hline
      1-1-1-2 & 0.500 & 0.268 \\\hline
      1-1-1-3 & 0.935 & 0.902 \\\hline
      1-1-2-2 & 0.500 & 0.293 \\\hline
      1-1-2-3 & 0.718 & 0.531 \\\hline    
      1-1-3-3 & 0.930 & 0.892 \\\hline    
%      1-2-2-2 & 0.500 & \cellcolor{gray}0.902 \\\hline    
      1-2-2-2 & 0.500 & 0.902 \\\hline    
      1-2-2-3 & 0.502 & 0.275 \\\hline
      1-2-3-3 & 0.613 & 0.308 \\\hline
      1-3-3-3 & 0.923 & 0.877 \\\hline
      2-2-2-2 & 0.500 & 0.531 \\\hline  
      2-2-2-3 & 0.501 & 0.308 \\\hline         
      2-2-3-3 & 0.502 & 0.276 \\\hline        
      2-3-3-3 & 0.713 & 0.533 \\\hline        
      3-3-3-3 & 0.919 & 0.868 \\\hline        
    \end{tabular}\caption {Average gate fidelity as a function of sequence of measurements when considering amplitude damping and dephasing channels. Numbers 1, 2, and 3 in the first column define the measurement time. For amplitude damping channel (1), is equivalent to $t=2\pi/d$, (2) is $t=3\pi/d$ and (3) is $t=4\pi/d$. For the phase damping channel (1) is equivalent to $t=\pi$, (2) is $t=3\pi/2$ and (3) is $t=2\pi$. The order of the numbers gives the time of each measurement on the first, second, third and fourth qubit.}\label{tab1}\end{table}\par}  %OK
    
    {\centering
    \begin{table}
        \begin{tabular}{|c|c|c|}
    \hline
      \textbf{Measurements} & \textbf{Fidelity of the} & \textbf{Fidelity of the} \\ 
             \textbf{$\;${Time}$\;$} & \textbf{AD Channel} & \textbf{PD Channel} \\ 

      \hline
      1-1-1 & 0.957 & 0.926 \\\hline
      1-1-2 & 0.250 & 0.002 \\\hline
      1-1-3 & 0.926 & 0.881 \\\hline      
      1-2-2 & 0.250 & 0.035 \\\hline    
      1-2-3 & 0.472& 0.080 \\\hline
      1-3-3 & 0.916 & 0.860 \\\hline
      2-2-2 & 0.250 & 0.035 \\\hline  
      2-2-3 & 0.255 & 0.080 \\\hline         
      2-3-3 & 0.485 & 0.860 \\\hline        
      3-3-3 & 0.915 & 0.860 \\\hline        
    \end{tabular} \caption{Resource gate fidelity as a function of sequence of measurements when considering amplitude damping and dephasing channels. The number 1, 2, and 3 in the first column define the measurement time. For amplitude damping channel (1) is equivalent to $t=2\pi/d$, (2) is $t=3\pi/d$ and (3) is $t=4\pi/d$. For the phase damping channel (1) is equivalent to $t=\pi$, (2) is $t=3\pi/2$ and (3) is $t=2\pi$. The order of the numbers gives the time of each measurement on the first, second, and third qubit.} \label{tab2}\end{table}\par} %OK
    
In the first column, the sequence of numbers represents the sequence of measurements and its respective time. For instance, in Table \ref{tab1}, for the amplitude damping, the parameter 1-1-1-1 in the first column means that all four necessary measurements to perform the gate occur on $t=2\pi/d$. To case 1-1-1-2, the first three measurements, on qubit 1, 2, and 3 occur on $t=2\pi/d$ and the fourth measurement on $t=3\pi/d$, to the case 1-2-2-3, the first measurement on qubit 1 occurs when $t=2\pi/d$, the subsequent measurements on qubit 2 and 3 occur when $t=3\pi/d$ and the fourth measurement occurs when $t=2\pi/d$ and so on to the rest of possible combinations. For the case of phase damping the idea is the same, only by changing the measurement times which are represented by the numbers 1, 2 and 3, now meaning $t=\pi$, $t=3\pi/2$, and $t=2\pi$, respectively. Note that the subsequent number is always bigger than the previous one, since the measurements need to be performed in order, from the first to the fourth qubit and, consequently, the measurement time over the next qubit can not be smaller. For the case of Table \ref{tab2}, the idea is analogous, but since only three measurements are required to create the resource state (on qubits 2, 3 and 4), each set of numbers is formed by three numbers. The first number indicates the time at which the second qubit is measured, the second number indicates the time at which the third qubit is measured, and the third number indicates the time at which the fourth qubit is measured. %OK 

Now, returning to our previous question, given the oscillatory behaviour of the cluster state fidelity in considering a highly non-Markovian environment, what can we expect from the gate fidelity in this scenario? Moreover, could delayed measurements result in better computational fidelity outcomes? Is it possible, with prior knowledge of the channel, to determine the best time to perform each measurement? Can non-unital channels, as amplitude damping, and unital channels, as phase damping, provide conceptually different results compared to the measurement times? To clarify these aspects, we analyze the average gate fidelity and the resource gate fidelity considering the non-Markovian amplitude damping and phase damping channels. %OK

A first important aspect to be emphasized is that the answer to these questions is independent of whether we are analyzing the $X$ or the $Z$ gate. Our results are equivalent for both cases. The reason for this equivalence is that both gates are performed by the same measurement sequence, with all qubits measured in the $\sigma_x$ basis. The difference on the $X$ and $Z$ gate, either in the computational process, which is presented in details in Section II, or the gate teleportation, presented in Section IV.A, is in the final correction given by $B_{\mathbf s}$ or $B_{\mathbf r}$, respectively. 
Despite the different corrections, they are unitary operations that in any case change the final fidelities, since the same operation done in two different states does not change the distance between them. 
Given this previous analysis, in order to better elucidate the differentiated aspects of the amplitude and phase damping channel, we study each of them separately below. %OK
%---------------------------------------------------------------------------------------
\subsection{MQBC under the action of the \\Amplitude Damping Channel }
%---------------------------------------------------------------------------------------
Analyzing Table \ref{tab1} (Table \ref{tab2}) we observe that the best results occur when the arrangement of the measurement times is coded by 1-1-1-1 (1-1-1), i.e. with all measurements performed in the first peak of the average cluster-state fidelity. On the other hand, the arrangement coded by 2-2-2-2 (or 2-2-2 to the case of resource gate fidelity), where all measurements are performed in the valley of the average cluster-fidelity, returns the worst. Indeed, for the amplitude channel, if any measurement is performed on the valley of the average cluster-fidelity both, the average gate fidelity and the resource gate fidelity, decrease significantly. Moreover, we note that the fidelity tends to be worse if the number of measurements in time which coincides with the valley of the average cluster-fidelity increases, and that the time of the last measurements, in the third or fourth qubit, is more crucial than the first ones. %OK
  
Also, analyzing Table \ref{tab1} and Table \ref{tab2} we observed that delayed measurements may actually result in better gate fidelities. The arrangement given by 3-3-3-3 (or 3-3-3), for instance, when all measurements are taken at the time that coincides with the second peak, is substantially greater than any other arrangement, except for those where the measurement times are replaced by the time which coincides with the first peak. These results show us that, to the case of highly non-Markovian amplitude damping channel, the best arrangement of the measurement times can be identified exclusively by means of the channel. Noting the expression of $p(t)$ on Eq. (\ref{ptad}) we see that the peaks of the average cluster-fidelity is given when $t=2n\pi/d$ where $n$ sets the order of the peak. In other words, once the channel is known, the position of the peak and valley is well characterized and, consequently, best arranged for the measurements. %OK

Indeed, the results to the gate fidelity, when considering the environment described by the amplitude damping, is expected. Getting greater gate fidelities when the measurement time coincides with the peaks time of the cluster-state fidelity, is an intuitive result. However, as we show next, the same is not necessarily true when considering the environment described by the phase damping channel. In this situation, since the channel is unital, counterintuitive results emerge. %OK
%---------------------------------------------------------------------------------------
\subsection{MQBC under the action of the \\ Phase Damping Channel }
%---------------------------------------------------------------------------------------
Analyzing Table \ref{tab1} and Table \ref{tab2}, as in the case where the MBQC is under the action of an AD channel, we observe that the best results occur when the measurement time arrangement is encoded by 1-1-1-1 (or 1-1-1 to the case of the resource gate fidelity). However, very counterintuitive results emerge when we carefully examine the average gate fidelity and the resource gate fidelity: depending on the number of measurements in time which coincides with the valley of the cluster state fidelity, the gate fidelity can indeed be high. Actually, to the case of average gate fidelity, with the exception of arrangement 1-1-1-1, one of the best results is obtained by arrangement 1-2-2-2, with three time measurements coinciding with the valley of the cluster state fidelity. Also, to the case of the resource gate fidelity, the arrangement defined by 2-3-3 gives equivalent results than that defined by 1-3-3 or 3-3-3. How can this be possible? What is behind this unexpected result? %OK

The main difference about the amplitude damping channel and the phase damping channel is that the latter is a unital channel. It means that in some circumstances the phase damping channel can act as a unitary operator or, at least, close to it. Inspecting the set of Kraus operator given by Eq. (\ref{epd1}) and Eq. (\ref{epd2}), it is straightforward to note that $E_1(t)\rightarrow 0$ and $E_2(t)\rightarrow \sigma_z$, where $\sigma_z$ is the well known Pauli $z$ matrix, when $L(t)\rightarrow -1$. It means that in this situation, the channel acts as a phase-flip and, in the case that $L(t) = -1$, the dynamics is unitary. Naturally, even in this situation, i.e. if the channel acts as a phase-flip gate, this is not sufficient to keep the fidelity equal $1$ but, as we will show here, it is a juncture of errors that fixes up the final operation. Let us first examine the function $L(t)$. %OK
\begin{figure}[h!]
\centering
\includegraphics[scale=0.3]{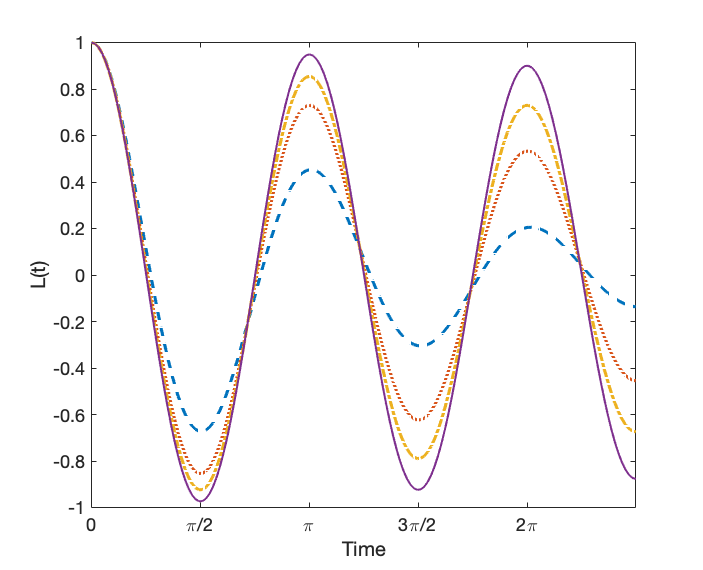}
\caption{$L(t)$ as a function of time. The solid curve (purple) represents $L(t)$ when $\tau=30$, while the dot-dashed (yellow), dotted (orange), and the dashed (blue) curves represent, respectively, the cases of $\tau=10$, $\tau=5$, and $\tau=2$.}\label{figLt}
\end{figure} %OK

Fig. (\ref{figLt}) shows the function $L(t)$ as a function of time for a set of different parameters $\tau$. As we can notice, as $\tau$ tends to increase, i.e. the degree of non-Markovianity grows, $L(t)$ tends to $1$ or $-1$ for some specific times.  On the one hand, for $t=\pi$ and $t=2\pi$, two of the preselected times where measurement is performed, $L(t)\approx 1$ and the phase damping Kraus operators tends to result in an identity operator. On the other hand, for $t=3\pi/2$, $L(t)\approx -0.92$ since that the phase damping Kraus operators incline to %OK
\begin{equation}\label{eap2}
    \begin{array}{*{20}{c}}
{E_1(t) \approx \left( {\begin{array}{*{20}{c}}
0.2&0\\
0&0.2
\end{array}} \right)}&{\rm and}&{E_2(t) \approx \left( {\begin{array}{*{20}{c}}
0.98&0\\
0&-0.98
\end{array}} \right)}
\end{array}.
\end{equation}
In this sense, it is natural to imagine that the set of measurement performed on $t=\pi$ and $t=2\pi$ could result in better gate fidelities, but a sequence of $\sigma_z$ operators can, in fact, also result in an identity. Indeed, the reason to reach high gate fidelity is that a specific sequence of phase-flips, plus a sequence of measurements on $\sigma_x$ basis, result in an equality given by %OK
\begin{equation}
P_{\mathbf s}[\rho_C(\psi_{in})]P_{\mathbf s} = P_{\mathbf s}[\tilde{\rho}_C(\psi_{in})]P_{\mathbf s}
\end{equation}
where $\tilde{\rho}_C(\psi_{in})\equiv[I\sigma_z^{(2)}\sigma_z^{(3)}\sigma_z^{(4)}\sigma_z^{(5)}{\rho}_C(\psi_{in})]$, with ${\rho}_C(\psi_{in})=|\Psi_C(\psi_{in})\rangle\langle \Psi_C(\psi_{in})|$ . It is important to note that a sequence of measurements, with the measurement time encoded by 1-2-2-2, results in a state similar $\tilde{\rho} _C(\psi_{in}) $ since (1) means a measurement on time $\pi$ (a time where the phase damping Kraus operators tends to result in an identity operator) and (2) a measurement on time $3\pi/2$ (a time where the phase damping Kraus operators tends to result in a $\sigma_z$ operator). %OK
In other words, when the computation is subjected to the phase damping channel, and the measurement time is given by the encoded sequence 1-2-2-2, $P_{\mathbf s}[\rho_C(\psi_{in})]P_{\mathbf s}\approx P_{\mathbf s}\mathcal{L}_t[\rho_C(\psi_{in})]P_{\mathbf s}$ with a gate fidelity given by $0.902$. %OK

\begin{figure}[h!]
\centering
\includegraphics[scale=0.3]{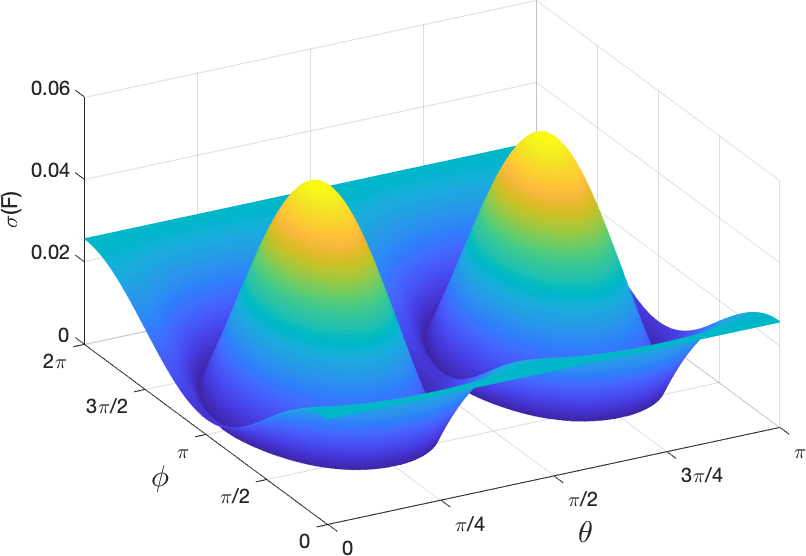}
\caption{Standard deviation of the average gate fidelity as a function of the initial state to the case of PD channel and measurement times represented by 1-2-2-2.} %OK
\label{fig3}
\end{figure}

Below, we make explicit the gate fidelity as a function of the second measurement time, considering the phase damping channel, to the case where $\tau=30$. We consider the time of first measurement equal to $\pi$ and the time of third and fourth measurement $3\pi/4$, so that the average gate fidelity immediately after the last measure is written as: %OK 
\begin{equation}
{F}_{pd}\approx \frac{1}{4}+\left(\frac{1}{4}+\frac{1}{8}{{\rm e}^{{-\frac {2\pi }{\tau}}}}\right) \left[ 1-\cos\left( 2t_2 \right) {{\rm e}^{-{\frac {2t_2+3\pi}{4\tau}}}}\right].
\end{equation}
It is important to note that the time of the second measure should be $\pi\le t_2\le3\pi/2$, since the minimum of the fidelity is reached when $t_2=\pi$ and the maximum when $t_2=3\pi/2$. This elucidates the fact that the gate fidelity is $0.293$ when the measurement time is encoded by 1-1-2-2, with the second measurement on time $\pi$, and increases to $0.902$ when the measurement time is encoded by 1-2-2-2, with the second measurement on time $3\pi/2$. %OK

Finally, another important aspect to emphasize, is the standard deviation of the average gate fidelity. Since we are considering an average, a relevant question is how fidelity is dispersed as a function of the initial state. To illustrate this aspect, we plot in Fig. (\ref{fig3}) the average gate fidelity as a function of $\theta$ and $\phi$, the two angles that define an arbitrary qubit initial condition, $|\psi_{in}\rangle = \cos\frac{\theta}{2}|0\rangle+\sin\frac{\theta}{2}{\rm e}^{i\phi}|1\rangle$. As we can notice, the standard deviation is indeed small, with a maximum value approximately equal to $0.055$ and average deviation equal to $0.019$. %OK
%=======================================================================================
\section{Conclusion}
%=======================================================================================
We study the dissipative dynamics of one qubit quantum gates in a measurement-based quantum computer when interacting with non-Markovian environments. We introduce the concept of average gate fidelity and show that it is identical to the $X$ and $Z$ gate. For highly non-Markovian environments, the average gate fidelity becomes extremely dependent on measurement times. We show that for the AD channel, the knowledge of the dissipative map is sufficient to determine the best measurement times, which is not necessarily true for the PD channel. For the case where each qubit is interacting with identical environments that introduce phase errors, we show that an error in one qubit can correct the error in another. This suggests that the construction of measurement-based quantum computers with identical qubits can be important for high fidelity computation. %OK
%=======================================================================================

\begin{acknowledgements}D. F. acknowledges support from Coordena\c{c}\~ao de Aperfei\c{c}oamento de Pessoal de N\'{i}vel Superior (CAPES), process number 88887.371735/2019-00. F.M. and F.F.F acknowledge support from Funda\c{c}\~{a}o de Amparo \`{a} Pesquisa
do Estado de S\~{a}o Paulo (FAPESP), project number 2019/00700-9 and 2019/05445-7, respectively.
\end{acknowledgements}

\newpage
\onecolumngrid
\begin{center}
\textbf{{\huge Appendix}}\\
\end{center}
$ $\\

%=====================================================================================
Here, we  present the analytical expressions of the $X$ and $Z$ operations, considering the  amplitude damping and phase damping channel.
Considering an arbitrary initial pure state $\left| {{\psi _{i}(\theta,\phi)}} \right\rangle  = \alpha(\theta)\left| 0 \right\rangle  + \beta(\theta,\phi)\left| 1 \right\rangle$, where $\alpha(\theta) = \cos \left( {\theta /2} \right)$ and $\beta(\theta,\phi) = \exp(+i\phi )\sin (\theta /2)$, such that $0 \le \theta  \le \pi$ and $0 \le \phi  < 2\pi$, the average gate fidelity as a function of measurement time is defined as:
\begin{equation}
   {F_{gate}}({t_1},{t_2},{t_3},{t_4}) = \frac{1}{\mathcal N}\sum\limits_{\theta,\phi}{F_m(\theta,\phi,t_1,t_2,t_3,t_4)}.
	\label{eq:media}
\end{equation}
where  $F_m(\theta,\phi,t_1,t_2,t_3,t_4)$ is the fidelity as a function of the initial state and the time of each one of the four measurements, considering the average over all 16 possible measurement results. Here, ${\mathcal N}$ is the normalization factor and can be define as ${\mathcal N}=\sum\limits_{\theta,\phi} \langle {{\psi _{i}(\theta,\phi)}}| {{\psi _{i}(\theta,\phi)}}\rangle$.% represents only the system fidelity for the cluster state dissipative dynamics (since no measurements are made on the system) or the average fidelity of the 16 possible measurement results for the operations $\pm X$ or $\pm Z$, in all cases to a single initial state $\theta$ and $\phi$, and has different expressions for each system as shown below.

For both $ X $ and $ Z$ operations, $F_m(\theta,\phi,t_1,t_2,t_3,t_4)\equiv F_m$ is equivalent and an analytical expression can be obtained cosidering the amplitude damping channel and the phase damping channel. For the amplitude damping it is given by:
%-------------------------------------------------------------------------------------
%\subsection{Amplitude damping}
%-------------------------------------------------------------------------------------
\begin{equation}
\begin{array}{l}
{F_m} = \frac{1}{2}{\rm{Re}}\bigg\{ \alpha {\left( \theta  \right)^2}\Big[ \alpha {\left( \theta  \right)^2}\left( {{p_4}\sqrt {{p_2}{p_4}}  + 1} \right) + \\
2{\left| {\beta \left( {\theta ,\phi } \right)} \right|^2}\left[ {\sqrt {{p_2}{p_4}} \left( {\sqrt {{p_1}{p_3}{p_4}}  - {p_4}} \right) + \sqrt {{p_1}{p_3}{p_4}}  + 1} \right] + \\
\left[ { - \beta {{\left( {\theta ,\phi } \right)}^2} - {{\left( {\beta {{\left( {\theta ,\phi } \right)}^*}} \right)}^2}} \right]\left( {\sqrt {{p_2}{p_4}}  - 1} \right)\sqrt {{p_1}{p_3}{p_4}} \Big]  + {\left| {\beta \left( {\theta ,\phi } \right)} \right|^4}\left( {{p_4}\sqrt {{p_2}{p_4}}  + 1} \right)\bigg\},
\end{array}
	\label{eq:xzad}
\end{equation}
where ${p_i} \equiv {p}({t_i})$, for $i=\{1,4\}$ where
\begin{equation}\label{ptad}
   p(t)= {e^{ - {\lambda}t}}{\left[ {\frac{{{\lambda}}}{{{d}}}\sin\!\left( {\frac{{{d}t}}{2}} \right) + \cos\!\left( {\frac{{{d}t}}{2}} \right)} \right]^2},
\end{equation}
with $d = \sqrt {2{\gamma_0}\lambda  - {\lambda^2}}$ and $\lambda$,$\gamma_0$ depending of the environment characteristic.

For the phase damping channel, 
%-------------------------------------------------------------------------------------
%\subsection{Phase damping} 
%-------------------------------------------------------------------------------------
\begin{equation}
\begin{array}{l}
{F_m} = \frac{1}{2}{\rm{Re}}\bigg\{ ({L_2}{L_4} + 1)\left( {\alpha {{\left( \theta  \right)}^4} + {{\left| {\beta \left( {\theta ,\phi } \right)} \right|}^4}} \right) + 2\alpha {\left( \theta  \right)^2}{\left| {\beta \left( {\theta ,\phi } \right)} \right|^2}\big[{L_2}{L_4}({L_1}{L_3}{L_4} - 1) + \\
{L_1}{L_3}{L_4} + 1\big] + \alpha {\left( \theta  \right)^2}{L_1}{L_3}{L_4}\left[ {\beta {{\left( {\theta ,\phi } \right)}^2} + {{\left( {\beta {{\left( {\theta ,\phi } \right)}^*}} \right)}^2}} \right](1 - {L_2}{L_4})\bigg\},
\end{array}
	\label{eq:xzdeph}
\end{equation}
%\begin{equation}
%\begin{array}{l}
%{F_m} = \frac{1}{2}{\rm{Re}}\{ {\alpha(\theta)^2}({\alpha(\theta)^2}({L_2}{L_4} + 1) + {\beta(\theta,\phi)^2}{L_1}{L_3}{L_5}(1 - {L_2}{L_4}) + \\
%{L_1}{L_3}{L_5}{({\beta(\theta,\phi)^*})^2}(1 - {L_2}{L_4}) + 2\beta(\theta,\phi){\beta(\theta,\phi)^*}({L_2}{L_4}({L_1}{L_3}{L_5} - 1) + {L_1}{L_3}{L_5} + 1)) + \\
%{\left| \beta(\theta,\phi) \right|^4}({L_2}{L_4} + 1)\} ,
%\end{array}
%	\label{eq:xzdeph}
%\end{equation}
where ${L_i} \equiv {L}({t_i})$, for $i=\{1,4\}$ where
\begin{equation}
    L(t) = {e^{ - t/2\tau }}\left[ {\frac{1}{u}\sin \left( {\frac{{ut}}{{2\tau }}} \right) + \cos \left( {\frac{{ut}}{{2\tau }}} \right)} \right],
	\label{ltpd}
\end{equation}
with $u = \sqrt {16{a^2}{\tau ^2} - 1}$, and $a$, $\tau$ depending of the environment characteristic.
%-------------------------------------------------------------------------------------
%=======================================================================================
\end{document}